\documentclass[sigconf]{acmart}
\usepackage{xurl}
\usepackage{graphicx}
\usepackage{float}
\AtBeginDocument{%
  }

\setcopyright{acmlicensed}
\copyrightyear{2026}
\acmYear{2026}

\acmConference[IndiaHCI '26]{17th Indian Conference on Human-Computer Interaction}
{October 29--31, 2026}{Hyderabad, Telangana, India}
\acmISBN{978-1-4503-XXXX-X/2018/06}

\begin{document}

\title{Translating Poetic Emotion into Dynamic Mathematical Visuals: A Strange Attractor-Based Approach}

\author{Subhadra Nayak}
\email{subhadran25@iitk.ac.in}
\orcid{0009-0007-4914-9972}
%\correspondingauthor
%\email{webmaster@marysville-ohio.com}
\affiliation{%
  \institution{HIVE Lab, Department of Design, Indian Institute of Technology - Kanpur}
  \city{Kanpur}
  \state{Uttar Pradesh}
  \country{India}
}

%\author{Lars Th{\o}rv{\"a}ld}
%\affiliation{%
%  \institution{The Th{\o}rv{\"a}ld Group}
%\email{larst@affiliation.org}
%\affiliation{%
%  \institution{Inria Paris-Rocquencourt}
%\affiliation{%
% \institution{Rajiv Gandhi University}
%\affiliation{%
%  \institution{Tsinghua University}
%\affiliation{%
%  \institution{Palmer Research Laboratories}
%\affiliation{%
%  \institution{The Th{\o}rv{\"a}ld Group}
%\email{jsmith@affiliation.org}
%\correspondingauthor
%\affiliation{%
%  \institution{The Kumquat Consortium}
%  \city{New York}
%\email{jpkumquat@consortium.net}

\renewcommand{\shortauthors}{Subhadra Nayak}


\begin{abstract}
Poetry conveys emotional experiences through language. This project investigates how poetic emotional experiences can be represented through dynamic mathematical visualisations while further developing a pipeline for this purpose. By leveraging a pretrained RoBERTa-based emotion classifier, the system analyses the fine-grained emotional states within a poem. These detected emotions are then mapped onto the parameters of strange attractors, complex structures from dynamical systems theory, to generate abstract, deterministic, and dynamic imagery. This work situates itself at the intersection of poetry, psychology, and mathematics, aiming to extend the poetic experience beyond language into the sensory domain.
\end{abstract}




\begin{teaserfigure}
 \includegraphics[width=\textwidth]{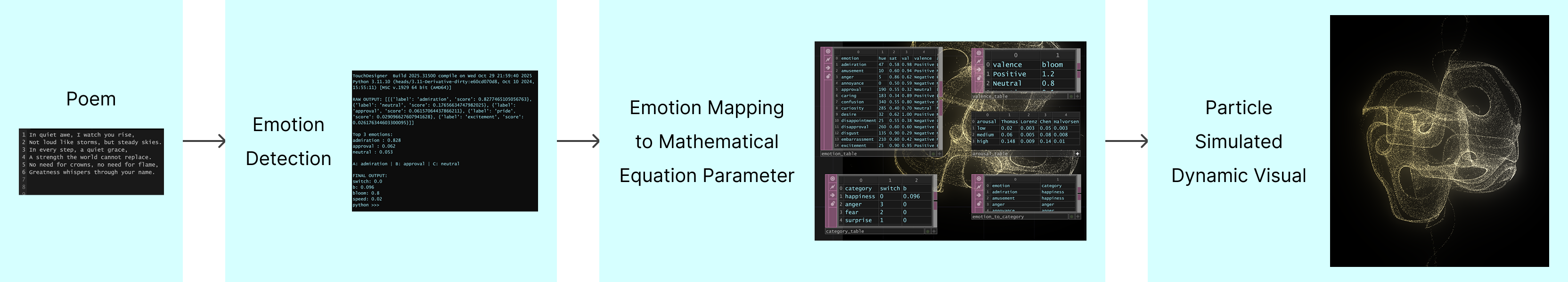}
  \caption{Overall pipeline of the proposed poem-to-visual system}
  \label{fig:teaser}
\end{teaserfigure}

\received{15 August 2026}
\received[revised]{01 October 2026}
\received[accepted]{10 October 2026}
\maketitle

\section{Introduction}
A poem is an arrangement of language that rewards its reader with a form of experience that is self-sufficient in itself to create a structured world \cite{vivas1954poem}. It is a form of writing that invites the kind of reading, attention, and analysis typically reserved for poetry, done in hopes of the meaningful possibilities of the language \cite{pierce2003defining}. A poem can have certain linguistic characteristics, such as meter, rhythm, metaphor, or imagery, but is not defined by the presence or absence of such a family of characteristics, but rather by the experience it provides.
 
Poetry can produce measurable emotional peaks, such as chills and goosebumps, with these responses tending to occur around poetic closures and being associated with socially and personally relevant content \cite{wassiliwizky2017emotional}.  Its content can allow readers to mentally simulate situations, people and experiences, while its formal and prosodic features can shape emotional responses through rhythm, meter, rhyme and anticipation. The readers' awareness of their experience and evaluation of the poem can contribute to an aesthetic emotional response \cite{johnson2022poetry}. The experience created by a poem evokes an emotional response, by both what it conveys and how it is constructed.

Poems often manifest through other art forms, such as performance/spoken word, music/audio, visual and audiovisual art (poetry films/installations), sculpture/visual poetry, and digital/media poetry, using kinetic text, manipulated voices, and other multimedia forms. This is done to produce an embodied experience and to make its silent properties perceptible, thereby expanding their spatiality beyond language \cite{psarras2018locating}. 

Mathematics and poetry are related in how patterns are formed and revealed. Poetry deals with patterns that reflect human experiences, largely influenced by emotions, while mathematics deals with patterns in the physical world, primarily governed by variables \cite{aharoni2014mathematics}. The proposed system aims to represent a poem's emotional content through dynamic visuals, primarily via a particle simulator. The main purpose is to extend the way poems are experienced through mathematics and to propose frameworks that could help achieve this. 

\section{Background and motivation}
Poems are often translated into new materials, spaces, bodies, imagery, sounds, and interfaces. Kinetic typography research employs time, motion, rhythm, tempo, sound, and spatial composition to enhance a poem's emotional impact \cite{kim2008enhancing}. The Muses of Poetry combines poetry with semantic analysis, real-time character animation, natural-voice interaction, facial expressions, and affective voice to control how a virtual character recites poem that expresses the emotional content of the poem \cite{sobiech2026and}. The Generative Visual Renku project combines poetry, generative art, animation, interface design, and computational semantics \cite{chow2010generative}. These approaches have made otherwise abstract textual features of the poem perceptible and increased its sensory range, making it more experiential.

The proposed system attempts to express expresses a poem's emotional content through the mathematics of a set of strange attractors that produce the dynamic visuals. The motivation for this approach comes from the ability of strange attractors to generate deterministic, dynamic, abstract visuals, making each simulation predictable.

\section{Research gaps and questions}

A substantial amount of work has been done to experience poetry in different forms. The previously mentioned research demonstrates several ways to visualise and analyse poems and their emotional content. However, the pipeline by which a poem's emotional content is used to create dynamic visuals governed by mathematical rules is not explored. The project develops an experimental prototype for this purpose and addresses the following research questions.

\begin{enumerate}

\item \textbf{RQ1:} Is there a way to visually embody a poem's emotional content such that it remains predictable on each simulation?
The answer was to control the visual parameter using a rule-based method, with a deterministic mathematical system to help. For this system, the chosen field of deterministic systems is strange attractors. Answering this RQ leads to another RQ,
\item \textbf{RQ2:} How can emotions expressed in a poem be mapped to the parameters of a mathematical system to construct such a visual representation?
\end{enumerate}

\section{Related Work}

\subsection{Emotion}

\subsubsection{Emotion and its classification}
Emotion is an intense, temporally unfolding experience that can arise from feelings, thoughts, memories, judgment, or imagination \cite{cabanac2002emotion}. There are different models that classify emotions based on different theories. Few examples as mentioned in \cite{sander2025theories} are: Discrete or basic-emotion models, Dimensional models, Appraisal models, Constructionist models, Componential models, Embodied and situated approaches. The models that have been worked upon are basic-emotion models and dimensional models, specifically valence-arousal, as discussed below.

\begin{enumerate}
\item \textbf{Paul Ekman's Basic Emotion Theory:} The theory proposes six basic emotions: happiness, sadness, fear, anger, surprise, and disgust. Later studies of facial expressions often included ``neutral'' as a baseline category rather than as a basic emotion.
\item \textbf{Valence-Arousal Model:} This represents emotional experience along two continuous dimensions: valence, which describes how pleasant or unpleasant an experience is, and arousal, which describes how activated or energetic the experience is. For example, excitement and fear can both involve high arousal, but excitement is generally pleasant whereas fear is unpleasant.
\end{enumerate}

Ekman's model focuses on identifying ``which emotion is present,'' whereas the valence-arousal model describes ``how pleasant and how activated the experience is.'' Together, these perspectives provide complementary ways of interpreting emotional content.

\subsubsection{Emotions Classifier used in the system}
The system uses a pretrained RoBERTa-based emotion classifier, \url{SamLowe/roberta-base-go_emotions}, accessed through the Hugging Face Inference API. The classifier was fine-tuned on the GoEmotions dataset which contains 58000 Reddit comments. 

There are 28 categories this model can recognise: admiration, amusement, anger, annoyance, approval, caring, confusion, curiosity, desire, disappointment, disapproval, disgust, embarrassment, excitement, fear, gratitude, grief, joy, love, nervousness, optimism, pride, realisation, relief, remorse, sadness, surprise and neutral

The classifier was selected primarily for its fine-grained emotion vocabulary for NLP, multi-emotion output, accessibility as a pretrained model, and suitability for exploratory prototyping. It was not selected because it is assumed to provide a ground-truth interpretation of poetic emotion. Given a poem as input, the classifier surfaces all detected emotions along with their weight scores.

\begin{figure}
\begin{center}
\includegraphics[scale=0.11]{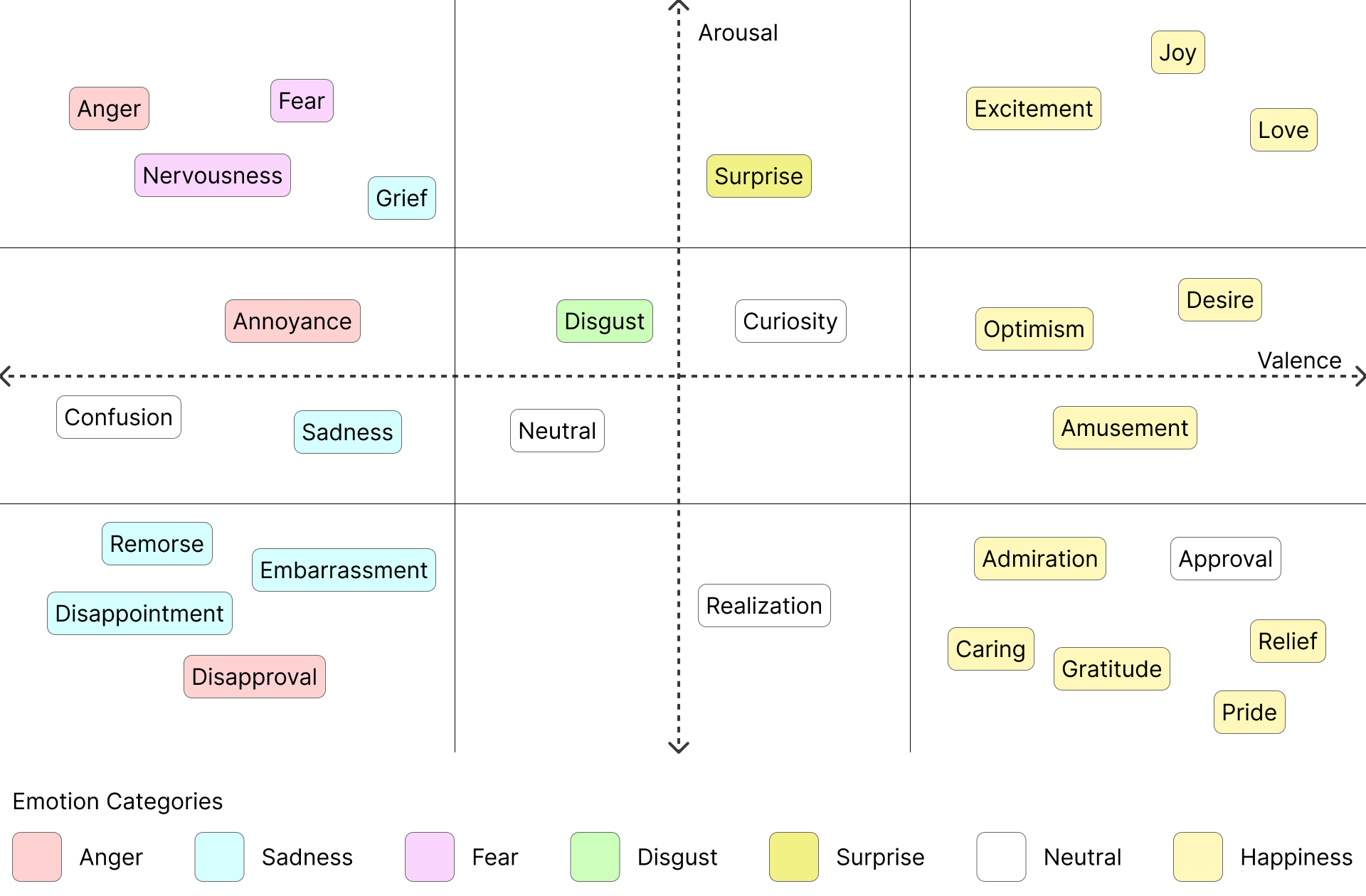} 
\caption{Valence–arousal representation of the 28 emotion labels used in the emotion-mapping framework.}
\label{fig_Mapping}
\end{center}
\end{figure}

\subsection{Mathematics}
Dynamical systems theory represents possible states in a state space and studies the trajectory traced through that space \cite{katok1995introduction}. A dynamical system is anything whose state changes over time according to rules. The \textit{state} is the information needed to describe the system at a given moment, and the rules determine how that state evolves. 

A chaotic system is a dynamical system whose behaviour is generated by definite rules but becomes difficult to predict over long periods because small differences in starting conditions grow rapidly. Two states that begin extremely close together can later follow very different trajectories \cite{ditto1995principles}. This is called sensitive dependence on initial conditions; the familiar \textit{butterfly effect}. The other features associated with chaotic system are deterministic, meaning the present state and the rules determine the next state and they are mostly non linear, meaning  effects are not proportional to causes as small changes can be amplified.

A strange attractor is an attracting invariant set in the phase space of a nonlinear dynamical system. Trajectories approaching the attractor remain confined to a bounded region and may show intricate, fractal-like geometry, aperiodic motion, and sensitivity toward initial conditions \cite{eckmann1985ergodic}. Its structure and dynamics depend on the governing equations, system parameters, and initial conditions.

In this work, four three-dimensional chaotic systems were used to generate distinct visual dynamics for the particle simulator: Thomas, Lorenz, Chen-Lee and Halvorsen systems. Changing the parameter values of these attractors can alter the resulting trajectory. In the proposed particle simulator, these attractors define the underlying visual dynamics, and their parameters are controlled by the poem's emotional content. This provides a connection between the representation of emotion by the mathematically based visual motion.

\section{Methodology}

The system translates the emotional content of a user-written poem into a dynamic visual representation generated through the mathematical behaviour of an attractor. The overall system consists of four interconnected stages:

\begin{enumerate}
    \item \textbf{Text input and emotion extraction}
    \item \textbf{Emotion-to-visual mapping}
    \item \textbf{Visual generation and its selection}
    \item \textbf{Visual rendering}
\end{enumerate}

The complete pipeline of the system can be expressed as shown in Figure \ref{fig_Pipeline}:

\begin{figure*}
\begin{center}
\includegraphics[width=\textwidth]{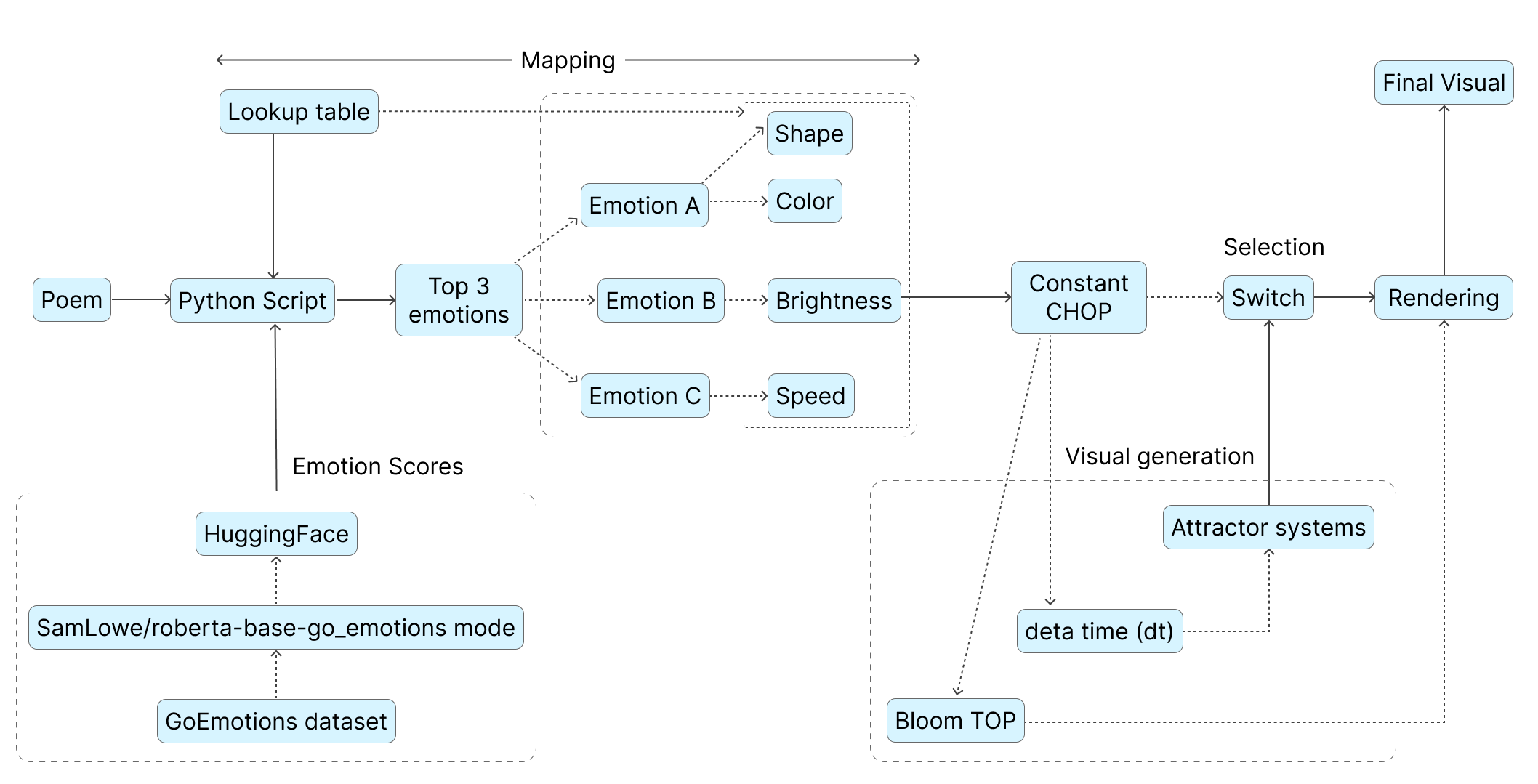} %\includegraphics[scale=0.25]{fig/Pipeline.png} 
\caption{Detailed system architecture of the proposed system, illustrating how emotions extracted from a poem are translated into dynamic visual representations.}
\label{fig_Pipeline}
\end{center}
\end{figure*}

%\begin{center}
%\textit{Poem}
%$\rightarrow$
%\textit{Emotion Scores}
%$\rightarrow$
%\textit{Top-3 Emotions}
%$\rightarrow$
%\textit{Emotion Mapping to Dynamical-System Parameters}
%$\rightarrow$
%\textit{Selected Visual}
%$\rightarrow$
%\textit{Rendering}
%\end{center}

\subsection{Text and Emotion Extraction}

The user enters a poem into the Text COMP in TouchDesigner, which is read by a Python script DAT. The Python script then retrieves emotion scores using the \texttt{SamLowe/roberta-base-go\_emotions} model from Hugging Face.

The model output is ranked according to the detected emotion scores, and the three highest-scoring emotions are selected. These three emotions are assigned distinct functional roles within the visual mapping pipeline as discussed in following section.

\subsection{Emotion-to-Visual Mapping Framework}

The Python script maps the three highest-scoring detected emotions to visual parameters using five lookup tables, collectively referred to as the emotion database:

\begin{itemize}

\item \texttt{emotion\_table}: Maps each emotion to hue, saturation, value, valence, and arousal. Thus, the table provides the colour representation of each emotion as well as its valence and arousal classifications. The valence-arousal categorisation used in the system follows a three-level representation. This can be seen in Figure \ref{fig_Mapping}.

\item \texttt{emotion\_to\_category}: Maps individual emotions to broader emotional categories based on Basic Emotion Theory. The categories used are happiness, sadness, fear, anger, surprise, disgust, and neutral.

\item \texttt{category\_table}: Maps the seven emotional categories to attractor parameters. Each category is associated with an attractor shape through the \texttt{switch} position, with Thomas = 0, Lorenz = 1, Chen-Lee = 2, and Halvorsen = 3. The parameter \texttt{b} is additionally used for the Thomas attractor to produce four distinct shapes.

\item \texttt{valence\_table}: Maps the three-level valence classification to the brightness or visual glow of the resulting shape through its bloom intensity.

\item \texttt{arousal\_table}: Maps the three-level arousal classification to particle movement parameters, specifically particle speed through the corresponding \texttt{dt} and offset values.

\end{itemize}

The three detected emotions are assigned different mapping paths:
\begin{enumerate}

\item \textbf{Dominant emotion A - shape and colour:} The dominant emotion first accesses the \texttt{emotion\_table}, which provides its predefined colour information in HSV format. The emotion is then passed through the \texttt{emotion\_to\_category} table to determine its broader emotional category. This category is subsequently used to access the \texttt{category\_table}, which determines the attractor selection through \texttt{switch} and provides the attractor-specific shape parameter \texttt{b}.

\item \textbf{Second emotion B - bloom:} The second-ranked emotion accesses the \texttt{emotion\_table}, from which its valence classification is obtained. The resulting valence (Positive, Neutral, or Negative) is then mapped through the \texttt{valence\_table}, which determines the corresponding bloom intensity.

\item \textbf{Third emotion C - movement:} The third-ranked emotion accesses the \texttt{emotion\_table} to obtain its arousal classification (Low, Medium, or High). This classification is then mapped through the \texttt{arousal\_table}, which determines the corresponding parameters governing particle movement.

\end{enumerate}

%\begin{itemize}
%
%\item \textbf{Dominant emotion A:} Shape and colour. The dominant emotion first accesses the \texttt{emotion\_table}, which provides its predefined colour information in HSV format. The same emotion is then passed through the \texttt{emotion\_to\_category} table to identify its broader emotional category. This category is subsequently looked up in the \texttt{category\_table}, which provides the attractor-selection value (\texttt{switch}) and the attractor-specific shape parameter \texttt{b}.
%
%\item \textbf{Second emotion B:} Bloom. The second emotion also first accesses the \texttt{emotion\_table}, but here only its valence classification is used. The resulting valence (Positive, Neutral, or Negative) is then looked up in the \texttt{valence\_table}, which specifies the corresponding bloom intensity.
%
%\item \textbf{Third emotion C:} Movement. The third emotion follows a similar two-step lookup. Its arousal classification (Low, Medium, or High) is obtained from the \texttt{emotion\_table}. This value is then used to access the corresponding row of the \texttt{arousal\_table}.
%
%\end{itemize}

%\begin{figure}
%\begin{center}
%\includegraphics[scale=0.11]{fig/Mapping.png} 
%\caption{Valence–arousal representation of the 28 emotion labels used in %the emotion-mapping framework.}
%\label{fig_Mapping}
%\end{center}
%\end{figure}

\subsection{Touchpoint of dependency connection}
The Python script then sends the mapped values into the seven channels of a constant CHOP. The constant CHOP, therefore, acts as a centralised parameter interface between the emotion-mapping layer and the visual-generation and rendering layers.
\begin{itemize}
\item \textbf{Switch:} Selects the attractor
\item \textbf{b:} Thomas attractor-specific; shape parameter
\item \textbf{Hue:} Controls the hue of the dominant emotion
\item \textbf{Saturation:} Controls the saturation of the dominant emotion
\item \textbf{Value:} Controls the value of the dominant emotion
\item \textbf{Bloom:} Controls bloom intensity in bloom TOP
\item \textbf{Speed:} Controls particle speed through attractor's dt and offset.
\end{itemize}

\subsection{Visual generation and its selection}
The dynamic visuals are generated by particles whose positions are determined by the previously discussed strange-attractor equations. These dynamical systems are implemented in TouchDesigner using GLSL and native TOP operators, with feedback used to retain and iteratively update the system state. All the attractors pass through the switch, whose value from the emotion mapping determines which attractor's output is passed forward.

\subsection{Visual Rendering}
\begin{figure*}
\begin{center}
\includegraphics[width=\textwidth]{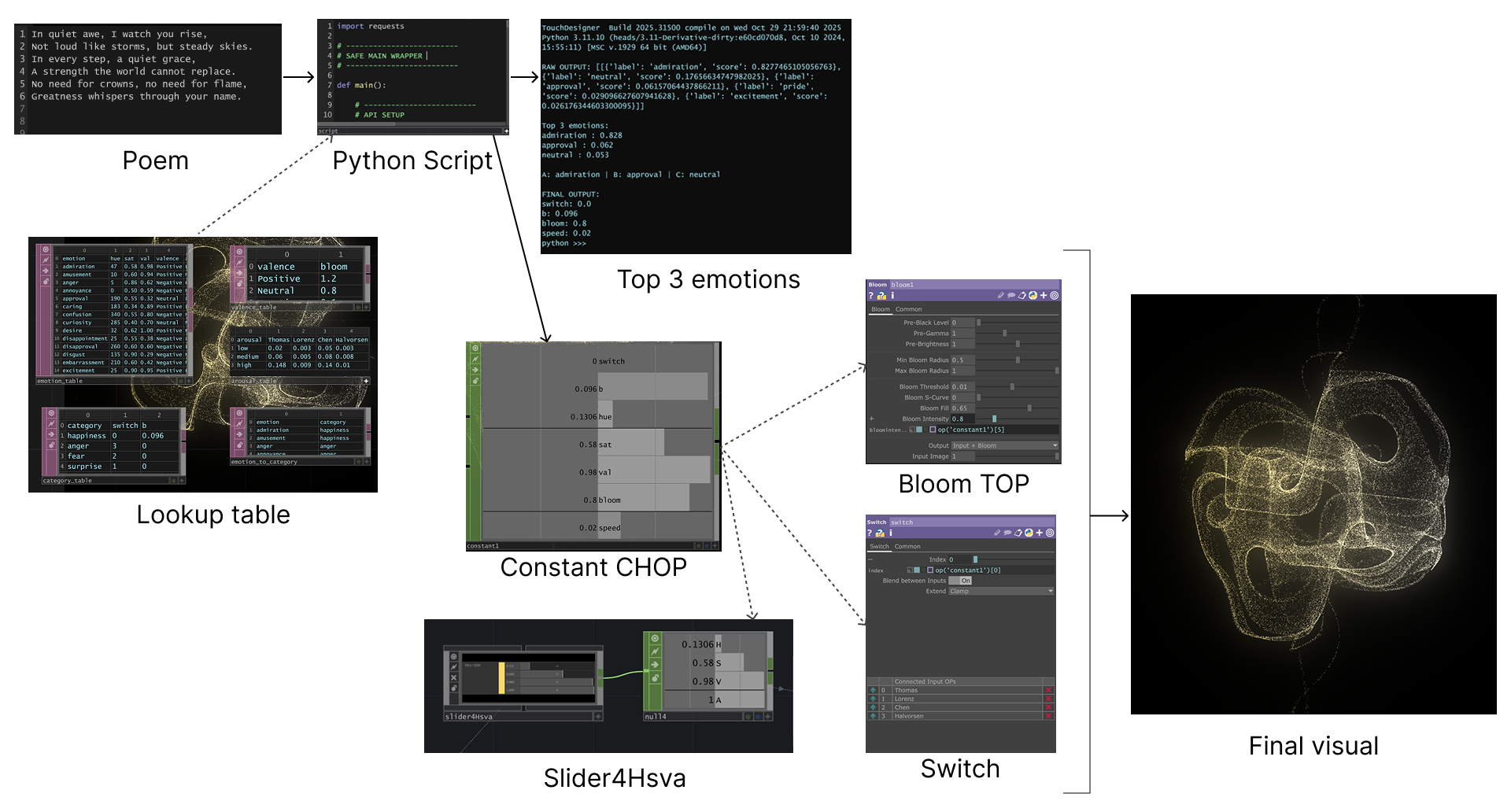} 
\caption{Implementation of the proposed poem-to-visual pipeline in TouchDesigner.}
\label{fig_Example}
\end{center}
\end{figure*}

Particle geometry is defined, then rendered using a camera and a light. After rendering, the system passes through the bloom TOP, and colour is adjusted by Slider4Hsva, which takes its value from the previously mentioned switch. These can be seen in Figure \ref{fig_Example}.

\section{Preliminary findings}
\begin{figure*}
\begin{center}
\includegraphics[width=\textwidth]{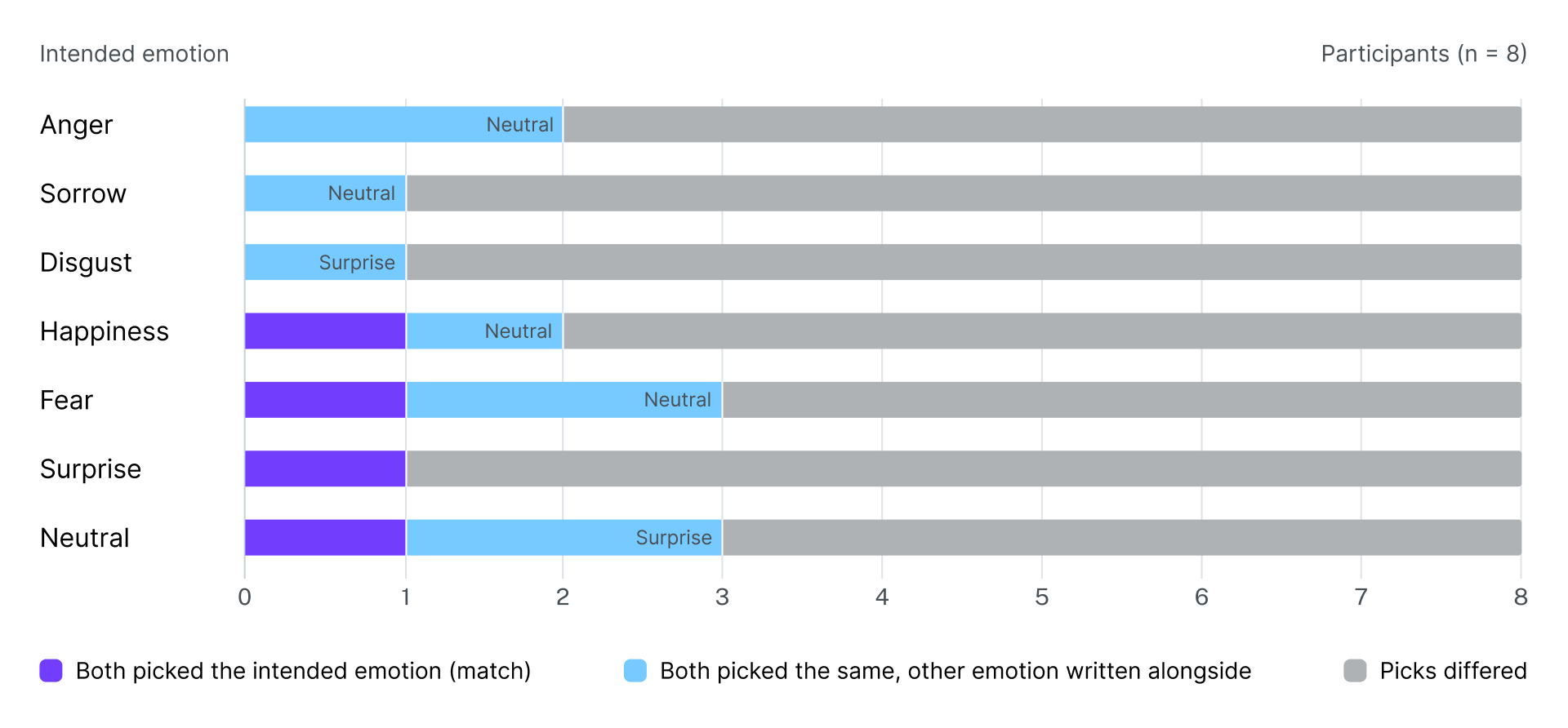} 
\caption{Analyzing participants' emotional choices for the poem and its visual representation, and assessing whether they align with each other.}
\label{fig_Testing}
\end{center}
\end{figure*}

A pilot study was conducted to simulate seven different categories of poems and create corresponding visuals for each. Eight participants were asked to describe the emotions they felt while reading each poem and viewing its related visual. This can be seen in Figure \ref{fig_Testing}. The visuals were presented in a different order from the poem, and the emotions available for selection were the same as those used to identify the shapes in the system: happiness, sadness, fear, anger, surprise, disgust, and neutral (labelled calm/neutral). Their participation was via a Google Form that included the poems and visuals linked to unlisted YouTube videos, which was sent to known people (convenience sampling) who might be interested in poetry and art, although reported engagement varied for both. Responses were analysed descriptively for exploratory purposes without using any statistical approach. 

Overall, agreement across poem-visual pairs was 23.21$\%$. For each of the seven emotions, evaluation was based on how each person responded to a poem and on how that response compared with the corresponding generated visual for the same poem. A match was considered when both their choices were the same. Of the 56 poem-visual pairs (8 participants multiplied by 7 emotions), 13 pairs matched, representing 23.21$\%$. It was noted that participants often selected "Neutral." 

\section{Discussion}
The pilot study was an exploratory, early-stage assessment with eight participants, aimed at identifying issues in the mapping rather than testing hypotheses or making generalisable claims. It cannot establish whether the system works or fails in general, and the findings may not generalise to other readers or viewers due to the small sample size. The pilot study also included a categorical response format, which may have limited participants' ability to fully describe their emotional experiences. An abstract visual can evoke a range of emotional qualities, such as calmness, sadness, or tension, which cannot be adequately represented by selecting a single category. As a result, some participants may have chosen "neutral" as a broad fallback option when the visual might not align with any specific category. A more open-ended and nuanced evaluation could provide a richer understanding of these responses.

The current approach of assigning parameters one-to-one may have its limitations. While the three highest-scoring emotions are derived from the poem, they are treated as separate, independent parameters. Consequently, the final visual is a combination of individually mapped emotional attributes rather than a holistic representation of the poem's emotional state. This separation may result in a visual representation that differs perceptually from the emotional character inferred from the poem, even when each mapping adheres to its established rule. This may be further work in this direction.

The emotion classification model introduces a potential source of variation within the system. During the development phase, there were instances in which some emotion that seemed prominent in a poem received a relatively low score from the classifier. For example, content related to "fear" might be assigned a lower score for "fear." Additionally, it was noticed that the model often leaned towards a "Neutral" classification. These variations can carry over into the mapping process and affect the resulting visual representation.

\section{Conclusion}
An experimental prototype of a system that connects a poem's emotional content to a mathematical equation has been created to visualise poetry. This study proposes a pipeline for this process. However, this is at an exploratory scale and opens future work around two areas. First, is the testing and analysis approach; second, the possibility of different mapping systems that could holistically represent a poem's emotional content. This could mean exploring how multiple emotional signals can be composed into a coherent visual representation. Both steps are cyclical to each other. These questions drive the need for deeper investigation into blended emotion-to-parameter mappings and more effective perceptual evaluation methods.

\bibliographystyle{ACM-Reference-Format}
\bibliography{sample-base}


\end{document}